\documentclass[12pt, preprint]{aastex}
\shorttitle{$Chandra$ Observations of AX~J1843.8$-$0352}
\shortauthors{Ueno et al.}

\begin{document}

\title{$Chandra$ Observations of a Non-Thermal Supernova Remnant Candidate 
AX~J1843.8$-$0352 and its Surroundings}

\author{Masaru Ueno, Aya Bamba, Katsuji Koyama}
\affil{Department of Physics, Graduate School of Science, Kyoto
University, Sakyo-ku, Kyoto, 606-8502, Japan
}
\email{masaru@cr.scphys.kyoto-u.ac.jp, 
bamba@cr.scphys.kyoto-u.ac.jp, koyama@cr.scphys.kyoto-u.ac.jp
}
\and
\author{Ken Ebisawa}
\affil{INTEGRAL Science Data Center
Chemin d'Ecogia 16 CH-1290 Versoix, Switzerland
}
\email{ebisawa@obs.unige.ch
}

\begin{abstract}
We present the {\it Chandra} results of AX~J1843.8$-$0352, a supernova 
remnant (SNR) recently identified with {\it ASCA}.
{\it Chandra} spatially resolved two components 
from this SNR: non-thermal and thermal ones. 
The morphology of the non-thermal component is clumpy and 
elliptical, elongated from the north to the south with 
a mean diameter of about 9$\arcmin$. The spectrum is 
fitted with a power-law model of photon index 2.1 and  
the east rim is associated with the non-thermal radio 
sources C and F \citep{Helfand1989}.
Therefore the non-thermal component is probably synchrotron X-rays
by energetic electrons accelerated at the shell of the SNR.
The thermal component is the brightest clump located within the 
non-thermal component and shows a spectrum of a thin plasma 
of about 0.7~keV temperature. 
Notable discovery is its peculiar morphology; a head of $50\arcsec 
\times 30\arcsec$ size near the south-east rim of the SNR 
and a $30\arcsec$-long ``jet'' pointing to the southwest.
Although this emission is associated with the west part of the radio 
source F, the absorption is twice larger than 
that of the non-thermal X-rays, or the bulk of the SNR emission.  
Therefore, it is unclear whether this peculiar plasma is a 
thermal component associated with AX~J1843.8$-$0352, a Galactic 
source located in the far side of our Galaxy, or an extragalactic 
source.
\end{abstract}


\maketitle

\section{Introduction}
Supernovae and their remnants (SNRs) play essential roles for the structure 
and evolution on the Galaxy; they are the major sources of galactic hot 
($10^{5}-10^{6}$~K) medium, their blast waves compress the interstellar 
medium and trigger successive star formations, and they produce and 
distribute heavy elements in the whole Galaxy and even in the 
intergalactic space.

SNRs would also be the most plausible birthplace of high-energy cosmic rays 
near to the knee energy ($\sim 10^{15.5}$ eV); the supporting evidence is
the detection of synchrotron X-rays from the shell of some of the SNRs 
(e.g.,\ Koyama et al.\ 1995, 1997; Slane et al.\ 2001), 
and the detection of TeV $\gamma$-rays from some of them 
(e.g.,\ Tanimori et al.\ 1998; Muraishi et al.\ 2000; Enomoto et al.\ 2002).   
Electrons should be accelerated to high energies up to about 1 TeV or 
even more, possibly by the first-order Fermi acceleration mechanism. 
Since the energy gain (acceleration) rates of electrons are 
proportional to $B$ (the strength of the magnetic field) and the energy loss 
(due to synchrotron radiation) rates are proportional to $B^{2}$, 
high energy electrons responsible 
for the synchrotron X-rays are likely to exist 
in a shell of rather weak magnetic field, where the radio flux 
should be faint (the flux is proportional to $B^{2}$).

SNRs may also be concerned as the major origin of 
the Galactic ridge X-rays (the GRXs). The GRXs are diffuse X-rays 
extending along the Galactic inner disk \citep{Koyama1986,Kaneda1997}. 
The spectrum shows thin thermal plasma of about 10$^{7-8}$~K, 
with prominent K-shell lines from highly ionized irons. 
There is mounting evidence that the GRXs are attributable 
to truly diffuse sources, not an integrated emission of many point sources 
\citep{Yamauchi1996,Ebisawa2001}. 
However, real origin of the diffuse GRXs has been 
a big puzzle; no known diffuse source can account for the observed spectrum 
and flux of the GRXs. Therefore the GRXs may predict presence of 
many diffuse X-ray sources, whether these are new X-ray SNRs or 
a new class of X-ray sources.  

Surveys with {\it ROSAT} and {\it ASCA} already found several 
new X-ray SNRs, which were later identified with radio faint SNRs. 
Much more sensitive {\it Chandra} obseravtion is expected to detect 
numerous such X-ray emitting SNRs, and resolve their spatial and 
spectral features. 
We have carried out a deep {\it Chandra} observations
on the Galactic plane region around AX~J1843.8$-$0352, 
a new SNR candidate identified with {\it ASCA} \citep{Bamba2001}. 
This paper reports on the nature 
of the diffuse structures and discusses possible implications 
on the origin of high-energy cosmic rays and 
the GRXs.
 
\section{Observations and Data Reduction}

The {\it Chandra} deep observations on the Galactic ridge were
performed on 2000 February 24--26 (here Observation 1 or Obs.\ 1) 
and on 2001 May 20--21 (here Observation 2 or Obs.\ 2). 
The targeted positions are 
R.A. =
$18^{\rm h}43^{\rm m}57\fs 8$, 
decl. =
$-04\arcdeg 04\arcmin 45\farcs 9$ (epoch 2000)
and R.A. =
$18^{\rm h}43^{\rm m}32\fs 1$, 
decl. =
$-03\arcdeg 54\arcmin 44\farcs 8$ (epoch 2000) 
for Obs.\ 1 and 2, respectively.
The observation regions were $17'\times17'$ fields of the ACIS-I arrays. 

The satellite and instrument are described by \citet{Weisskopf1996} 
and \citet{Garmire1997},
respectively. AX~J1843.8$-$0352 
lies near the northwest edge and the east part of the ACIS-I array, 
for Obs.\ 1 and 2. 
Data acquisition from ACIS-I was made in Timed-Exposure Faint mode 
with the chip readout times of 3.24~s. Data reduction and 
analysis were made using the {\it Chandra} Interactive Analysis of 
Observations (CIAO) software version 2.2. 
Using Level 2 processed events provided by the pipeline
processing at the {\it Chandra} X-ray Center, we selected 
the {\it ASCA} grades\footnote{see
http://asc.harvard.edu/udocs/docs/POG/MPOG/index.html} 0, 2, 3, 4 
and 6, as X-ray events.  The other events due to charged
particles and/or hot and flickering pixels were removed. 
We also removed high background data in the Obs.\ 1 
where total count rates were larger than $5.5$~counts~s$^{-1}$.
The effective exposures were then about 94~ks and 99~ks for 
Obs.\ 1 and 2, respectively.

\section{Analysis and Results}

\subsection{Over-all Morphology}

In the raw images obtained by these observations, we see numerous point 
sources.  We at first picked up point sources 
from the raw images in the energy bands of 
0.5$-$3.0~keV, 3.0$-$8.0~keV, and 0.5$-$8.0~keV 
using the program WAVDETECT \citep{Freeman2002} with the 
significance criterion at 4.0$\sigma$, which corresponds to about 
6 -- 33~counts~s$^{-1}$, 
depending on the size of the point-spread function (PSF), 
the energy band, and the background level.
Then 274 sources are found in the two observations, of which 
225, 116, and 271
are detected in the energy bands of 
0.5$-$3.0~keV, 3.0$-$8.0~keV, and 0.5$-$8.0~keV, respectively. 
Surface density of the point sources detected in the 3.0$-$8.0~keV 
band is not significantly higher than that expected from the 
extragalactic sources seen through the Galactic plane.  Therefore, 
most of the hard X-ray sources are considered to be extragalactic 
\citep{Ebisawa2001}.  On the other hand, soft X-ray point 
sources show significant excess over the extragalactic population. 
They have low temperature ($< 1$~keV) thermal spectra and some of them 
exhibit flare-like temporal variations \citep{Ebisawa2002}. 
Therefore, most of these soft X-ray sources are presumably nearby active 
stars.

The combined image (Obs. 1 + 2), after the exposure correction and 
smoothing with a Gaussian kernel of $\sigma$ = 6$\arcsec$ 
is shown in Figure~1, 
where the point sources of $>5.0 \sigma$ detection in the 
0.5$-$8.0~keV band are given by crosses.
In Figure~1, we can see an extended complex at the north 
(shown by the solid-line ellipse). 
Since the point sources are scattered uniformly over the field of view, 
the extended X-ray structure is unlikely to be a local enhancement 
of many point sources, but will be a really diffuse source. 
The center position of the diffuse source is at around 
R.A. = 18$^{\rm h}$43$^{\rm m}$50$^{\rm s}$, 
decl. = $-03\arcdeg 52\arcmin 00\arcsec$ (epoch 2000) 
and the emission is extended in the elliptical shape 
of $11\arcmin \times 7\farcm 5$.
The center position and morphology correspond to the {\it ASCA} source 
AX~J1843.8$-$0352 \citep{Bamba2001}, 
hence we use the {\it ASCA} name, hereafter. 

We found that AX~J1843.8$-$0352 consists of numerous clumps.  
We name the brightest spherical clump in the south-eastern 
part of AX~J1843.8$-$0352 as CXO~J184357$-$035441 from its 
central position. 
A closed-up view of this clump in the 1.0--6.0~keV band is 
given in Figure~2, together with the 20~cm (1.5 GHz) contour of 
a radio source F \citep{Helfand1989}. 
A notable feature is its peculiar shape: an elliptical head 
with a jet-like tail of $30\arcsec$-long to the southeast.   

\subsection{AX~J1843.8$-$0352}

The X-ray spectra of AX~J1843.8$-$0352 were separately made 
for Obs.\ 1 and 2, from the elliptical region of 
$11\arcmin \times 7\farcm 5$ given by the solid line in Figure~1. 
An elliptical region surrounding CXO~J184357$-$035441 
(dotted line in Figure~2) was excluded to extract the energy spectrum. 
Although Obs.\ 2 covers full  region, Obs.\ 1 covers only about 1/3 
at the southern part. The point source events,
which are in the PSF (90\% encircled radii) circles 
around the point sources detected with the WAVDETECT software 
(see subsection 3.1), were removed from the source events. 
The background regions were selected at the same Galactic latitude 
as the source region, so that the GRXs are properly eliminated
(e.g., Kaneda et al.\ 1997). 
The background regions for Obs.\ 1 and 2 are shown 
with dotted circles in Figure~1. 
The point source contribution was excluded from the background spectra, 
in the similar way as the source regions. 

The background-subtracted spectra for Obs.\ 1 and 2 have no emission 
line as shown in Figure~3. We fitted the spectra with a power-law model. 
The fits are acceptable with 
$\chi ^{2}/$degree of freedom (d.o.f.) $= 18.9/27$ and $48.1/52$ 
for Obs.\ 1 and Obs.\ 2, respectively, and 
the best-fit parameters are given in Table~1.
Although the spectrum of Obs.\ 1 represents only the 
southern part of AX~J1843.8$-$0352, the spectral parameters except the flux 
are consistent with those of  Obs.\ 2, the full spectrum of 
AX~J1843.8$-$0352.
We therefore carried out a simultaneous fit to the  Obs.\ 1 and 2 spectra
with a power-law model leaving each normalization of the two observations 
independent. The fit is acceptable with the best-fit parameters 
given in Table~1. 

To examine the possibility for a thermal plasma origin, 
we also proceeded a simultaneous fit with a model of a thin thermal plasma 
in non-equilibrium ionization, an NEI model (Borkowski et al.\ 2001; 
a revised version from an original code by \citet{Hamilton1984}). 
This NEI model is acceptable with the best-fit parameters in Table~1.
The metal abundances, however, is lower ($\sim 0.2$~solar)
and the temperature is higher ($\sim$~5.4~keV) than any of the diffuse 
thermal sources (e.g., SNRs and star forming regions) in the Galaxy.

The spatial correlation between the diffuse X-ray and radio
is demonstrated by the 20~cm VLA contours in Figure~1, with 
the designation of the radio source (A--I) given by 
\citet{Helfand1989}.  
The X-ray emission consists of many clumps  filling  an ellipse
which is partly outlined by the radio sources C and F, the SNR candidates
proposed by \citet{Helfand1989}. The south tail of the radio source G
is also correlated with the X-ray enhancement. Though the source H is said 
to be extragalactic because of its infrared emission \citep{Helfand1989}, 
it is also inside AX~J1843.8$-$0352 and may be associated. 
Note, however, no significant X-ray emission is associated with the source E, 
which is a partial radio shell and an SNR candidate suggested 
by \citet{Helfand1989}.

\subsection{CXO~J184357$-$035441}

Since the WAVDETECT software detected four point sources 
in CXO~J184357$-$035441 (the crosses in Figure~2), 
we examined in detail whether these sources are really point 
sources or not, and found no clear evidence for point sources.  
The brightest source located at south-west, for example, has 
a wider radial profile than PSF as is demonstrated in Figure~4.  
The light curves of all these ``point sources'' show 
no time variability on the time scale of 10$^{3}$ to 10$^{5}$~s. 
Although statistics are limited, the X-ray spectra are found 
to be consistent with that of the whole of CXO~J184357$-$035441.  
Furthermore the total flux of these ``point sources'' is 
only 1/4 of that from CXO~J184357$-$035441. 
Accordingly, we regard the ``point sources'' are due to 
local enhancement of the bright diffuse source 
CXO~J184357$-$035441. 

The X-ray spectrum of CXO~J184357$-$035441 was made from 
the full  elliptical region of $50\arcsec \times 30\arcsec$ 
shown in Figure~2. We used the same background spectrum as 
that for AX~J1843.8$-$0352 in Obs.\ 2. 
As is given in Figure~5, the background-subtracted spectrum 
is much softer than that of AX~J1843.8$-$0352, and shows 
clear emission lines at 1.85 and 2.41~keV, which are equal to 
or slightly lower than those from He-like Si and S. 
We fitted the spectrum with an NEI model. 
This model is acceptable ($\chi^{2}$/d.o.f = 22.2/39) with 
the best-fit parameters given 
in Table~2 and the best-fit model in Figure~5.

\section{Discussion}

\subsection{AX~J1843.8$-$0352}

We found that the newly identified SNR AX~J1843.8$-$0352 
has many X-ray clumps, which are globally filling an elliptical region. 
The X-ray spectrum is well-fitted with either a power-law 
or a thin thermal plasma model, 
similar to the {\it ASCA} spectrum \citep{Bamba2001}.
{\it Chandra} gives more severe constraint on the thin thermal 
parameters than {\it ASCA}, due mainly to the removal 
of the contamination from the thermal source CXO~J184357$-$035441.
The thermal  scenario of AX~J1843.8$-$0352 requires uncomfortably 
high temperature ($>3.8$~keV) and low metal abundance ($<0.34$~solar) 
(Table~1). This strengthens the conclusion made with 
{\it ASCA} \citep{Bamba2001} that the X-rays of AX~J1843.8$-$0352 is 
non-thermal origin with a power-law spectrum.

The absorption ($N_{\rm H}$) of $(3.2-4.5) \times 10^{22}$~cm$^{-2}$ 
is slightly larger than, but roughly similar to the {\it ASCA} 
result. We therefore adopt the source distance to be 7~kpc following 
the discussion of \citet{Bamba2001}. 
The X-ray luminosity (2.0--10.0~keV) and the source size are then 
estimated to be $1.5\times 10^{34}$~erg~s$^{-1}$ and 18~pc 
(mean diameter), respectively.  

Since the spectrum is a power-law with the photon index of 2.1 
(the energy index of 1.1), 
the X-rays would be due to synchrotron emission from high energy 
electrons in a power-law distribution. The corresponding spectral index 
of the electrons is 3.2, which is larger than the value of 2 which is 
expected by the first-order Fermi acceleration. 
This steepening in the X-ray band is also seen in all of the well 
established SNRs as a site of high energy electrons 
(SN~1006, G347.3$-$0.5, and RX~J0852.0$-$4622, 
e.g., Koyama et al.\ 1995, 1997; Slane et al.\ 2001), and their photon 
indices are very similar (between $2.2-2.6$). 

For a wide-band study of the synchrotron emission, we fitted 
the X-ray spectrum with an SRCUT model \citep{Reynolds1998}, assuming 
the energy index ($\alpha$) in the radio band (1~GHz) to be 0.5 or 0.6. 
The derived cut-off energy and the flux density at 1.5~GHz are shown 
in Table~3. Here, the absorption is consistent with the fit with 
the power-law model. 
Both of the predicted radio fluxes at 20~cm by the SRCUT model 
(for $\alpha= 0.5$ and $\alpha = 0.6$) are  
smaller than the summed flux of the radio clumps C, F, G and H 
($\sim 1$~Jy; Helfand et al.\ 1989). 
This apparent inconsistency may be due to our simple assumption
that magnetic field strength ($B$) and density of high energy electrons 
($n_{\rm e}$) are uniform in the ellipse. 
More realistic assumption is that the radio clumps 
have higher $B$ than the average, 
hence produces high contrast of the radio flux. 
On the other hand, X-ray flux contrast is small, 
because $n_{\rm e}$ of higher energy electrons
(which emits X-rays) are smaller than the average, due 
to the  high synchrotron energy loss at high $B$. 
This scenario is consistent with the observed X-ray and radio 
morphology. 
In this case, the SRCUT model predicts that in those regions 
the cut-off energy is lower and the X-ray spectrum is steeper than 
the other part. The limited photon flux, however, does not allow us to 
verify this scenario. 
We note that G347.3$-$0.5, which is one of the well established SNRs as 
the site of high energy electrons, also shows no good correlation 
with the radio and X-rays \citep{Slane1999}. The primary candidate for 
the reason is the variation of the strength of magnetic field ($B$) 
from place to place. To clarify this issue, 
fine spatial and deep observations 
in both the radio and X-ray bands are encouraged.

Assuming energy equi-partition \citep{Pacho1970}, we 
derived the strength of magnetic field $B_{\rm eq}$ and 
the total energy of non-thermal electron $E_{\rm eq}$ in Table~3. 
The total energy of non-thermal electron is about 1/20 
times that of SN~1006 \citep{Dyer2001}. The maximum energy of 
electrons is estimated to be  140 ($\alpha$  =0.5) or 
200~TeV ($\alpha$ = 0.6).  The synchrotron cooling times of 
the electrons are $\sim 1400-1700$~yr.
 
\subsection{CXO~J184357$-$035441} 

The X-ray energy spectrum of CXO~J184357$-$035441,
thin thermal spectrum with 0.7~keV temperature and 
solar abundance, is typical for young and intermediate-age SNRs. 
Since the projected position of this source is 
in the SNR AX~J1843.8$-$0352 and is associated with the non-thermal 
radio source F \citep{Helfand1989}, a naive scenario is that 
CXO~J184357$-$035441  is a thermal counterpart of the SNR. It may 
be possible that we have a large shell whose soft X-ray emission is 
invisible because heavily absorbed. While a reverse shock 
is propagating toward the center, the core region makes 
harder spectrum. Hence, CXO~J184357$-$035441 might be one of the 
clumps made by the reverse shock. Adopting the same distance 
as AX~J1843.8$-$0352 (7 kpc), we estimate that 
the X-ray luminosity (0.7--10.0~keV) is $2.7\times 10^{35}$~erg~s$^{-1}$ 
and the  physical size is  $1.7 \times 1.0$~pc of ``head'' and 
1.0~pc-long ``jet''. Assuming a uniform density of a prorate sphere 
with $1.7 \times 1.0 \times 1.0$ pc diameters, 
the density of the ``head'' is  $\sim 15$ cm$^{-3}$. 
From the ionization parameter ($n_{\rm e}t$) of 
$(1.4 - 83)\times 10^{10}$~cm$^{-3}$~s, 
this plasma would  be heated-up very recently ($30 -1700$ years ago),
possibly by the reverse shock. 


Since  CXO~J184357$-$035441 exhibits twice larger absorption
than AX~J1843.8$-$0352, this source is likely located at the distance 
of 14 kpc. Derived plasma luminosity, size and density are still 
consistent with being a young SNR behind AX~J1843.8$-$0352.

Since the absorption of CXO~J184357$-$035441 is nearly the same
as the total Galactic absorption through the Galactic plane 
\citep{Ebisawa2001}, a possibility of an extragalactic source
can not be excluded. The jet-like morphology is also unusual 
for an SNR plasma, but rather resembles the hot plasmas associated 
with the jet sources, like SS433 or AGN. 
It is unclear whether CXO~J184357$-$035441 is a thermal component 
associated with AX~J1843.8$-$0352, another SNR located in the
far side of our Galaxy, or an extragalactic source.

\subsection{Implications for the GRXs}

The present discovery of many clumps with low surface brightness 
both in the hard (AX~J1843.8$-$0352) 
and soft (CXO~J184357$-$035441) bands may have important hints 
on the GRXs. The GRXs have been found to exhibit three components,
 with two thermal plasma and hard non-thermal-tail
\citep{Kaneda1997, Yamasaki1997}.    
The high  temperature plasma (7--10~keV) emits strong K-shell lines 
from He-like irons (the hard component), while low temperature
plasma (0.7~keV) has 
prominent K-shell lines from He-like Si and S, 
which is very similar to CXO~J184357$-$035441. 
The GRXs are prominent in the inner disk 
of the Galactic longitude $b = \pm 30^{\circ}$, and the positions of 
AX~J1843.8$-$0352 and CXO~J184357$-$035441 are near the edge of this 
inner disk. 

The absorption corrected surface brightness of AX~J1843.8$-$0352 
in the 2.0$-$10~keV band is 
$2.5 \times 10 ^{-7}$~erg~cm$^{-2}$~s$^{-1}$~sr$^{-1}$, 
which corresponds to $\sim 2.4$ times that of the hard component
(high temperature plasma and non-thermal tail)  of GRXs 
at b$\simeq$0$\arcdeg$ \citep{Kaneda1997,Yamasaki1997}. 
The size of AX~J1843.8$-$0352 ($\sim$65~arcmin$^{2}$) 
is $\sim 1/8$ times the region covered by Obs. 1 and Obs. 2 
($\sim$510~arcmin$^{2}$).
From these values, we can estimate that in the 2.0$-$10~keV band, 
the contribution of AX~J1843.8$-$0352 is $\sim$1/3 of the hard 
component of the GRXs from the region covered by the $Chandra$ observations. 

We also estimated the contibution of Si and S K-line emission from 
CXO~J184357$-$035441 into the GRXs.
The flux of Si and S K-lines from CXO~J184357$-$035441 are 
$7.4 \times 10^{-14}$ and $4.2 \times 10^{-14}$~erg~cm$^{-2}$~s$^{-1}$, 
respectively. 
On the other hand, since the brightness of Si and S K-lines of the GRXs are 
$6.8\times 10^{-9}$ and 
$7.8\times 10^{-9}$~erg~cm$^{-2}$~s$^{-1}$~sr$^{-1}$, respectively 
(cf.\ Kaneda et al.\ 1997), the fluxes of these lines from the region 
observed with $Chandra$ are expected to be $2.9\times 10^{-13}$ and
$3.4 \times 10^{-13}$~erg~cm$^{-2}$~s$^{-1}$. 
Therefore, to explain the Si and S K-line components of the GRXs 
with sources like CXO~J184357$-$035441, $\sim 4-8$ such sources are 
needed in each two $Chandra$ observations. 
In order to see whether sources of new population like 
CXO~J184357$-$035441 are omnipresent or not, we encourage to perform 
further deep-exposure observations on the Galactic disk with $Chandra$ 
or $XMM$-$Newton$.

\section{Summary}

1.  We confirmed the diffuse hard X-ray emission, AX~J1843.8$-$0352, 
    which is a recently identified SNR with $ASCA$ 
    in a non-thermal radio complex.
	
2.  The X-ray spectrum is represented with 
    a non-thermal power-law model of photon index $\sim$2.1. 
    Together with the elliptical morphology with a mean diameter of 18~pc, 
    we confirm that AX~J1843.8$-$0352 is 
    an SNR which predominantly emits synchrotron X-rays.

3.  We discovered a new X-ray source CXO~J184357$-$035441 within 
    AX~J1843.8$-$0352. CXO~J184357$-$035441 is extended (several arcsec) 
    and exhibits thin-thermal spectrum. Its morphology is peculiar with 
    an elliptical head and a jet-like tail.

4.  The hydrogen column density toward CXO~J184357$-$035441 is 
    $\sim 7\times 10^{22}$~cm$^{-2}$, 
    which is nearly twice larger than that of AX~J1843.8$-$0352. 
    Whether CXO~J184357$-$035441 is a part of the SNR or 
    a background source is debatable.

\acknowledgments
The authors thank Shigeo Yamauchi for fruitful discussions. 
We are grateful to all the members of the {\it ASCA} Galactic 
Plane Survey Team.
M.U. and A.B. are supported by JSPS Research Fellowship for Young 
Scientists.

\clearpage

\begin{figure} 
\plotone{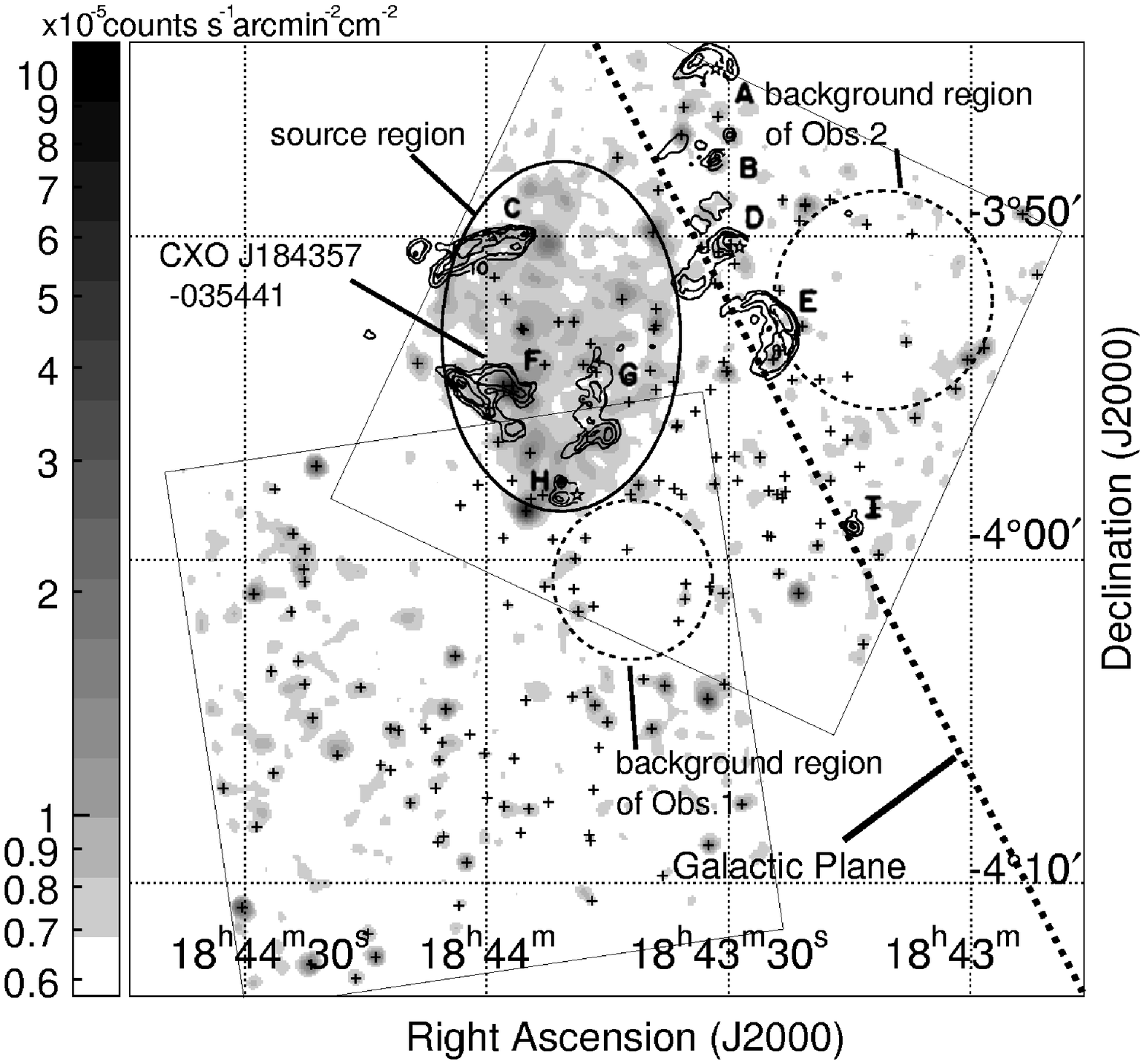}
\figcaption{
The combined ACIS-I image of Obs.\ 1 and Obs.\ 2 in the 1.5--8.0~keV 
band overlaid on the VLA 20~cm radio contours by \citet{Helfand1989}. 
The X-ray image is corrected for the exposure, smoothed with a Gaussian 
kernel of $\sigma$ = 6$\arcsec$ and plotted from 
$5.7\times 10^{-6}$ to 
$1.1\times 10^{-4}$~counts~s$^{-1}$~arcmin$^{-2}$~cm$^{-2}$ 
in logarithmic scale.
The squares of the narrow solid lines show the FOVs of Obs.\ 1 and 2. 
The source and background regions for AX~J1843.8$-$0352 are 
designated by the thick solid line and broken lines, respectively.
The point sources detected by the WAVDETECT software 
with significance $>5.0\sigma$ are designated by the crosses. 
Radio sources are labeled following \citet{Helfand1989}.}
\end{figure}

\begin{figure} 
\plotone{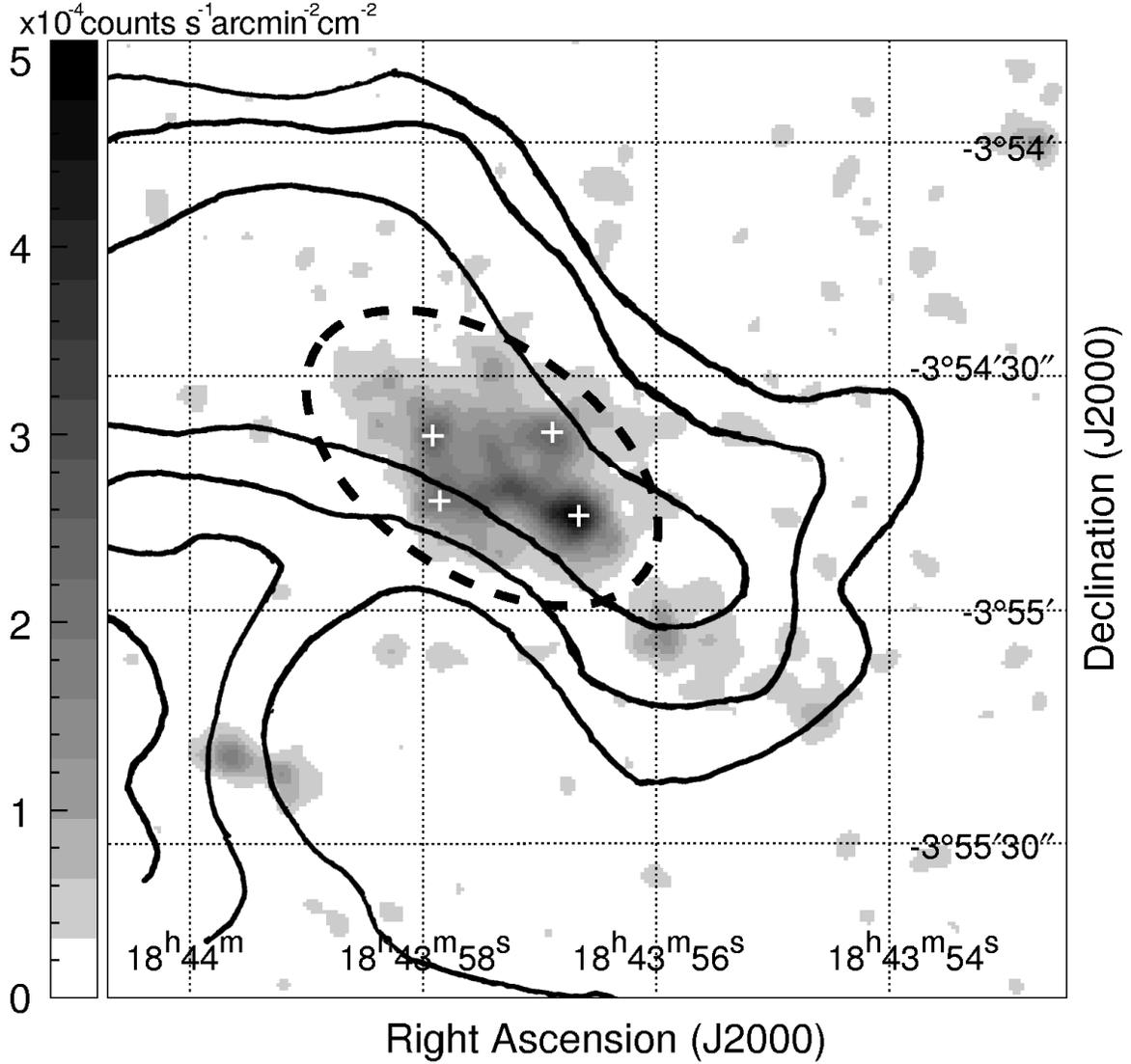}
\figcaption{The closed-up view of CXO~J184357$-$035441 in 
the energy band of 1.0--6.0~keV. The image is smoothed with 
a Gaussian kernel of $\sigma$ = 1$\farcs$5  and plotted from 
0 to $5.1\times 10^{-4}$~counts~s$^{-1}$~arcmin$^{-2}$~cm$^{-2}$ 
in linear scale.
The 20~cm-radio continuum source F \citep{Helfand1989} is given  
by solid-line contours. 
The ellipse region for extracting the spectrum of 
CXO~J184357$-$035441 is shown by the dashed line. 
The four ``point sources'' detected in
CXO~J184357$-$035441 are designated by the white crosses.
}
\end{figure}

\begin{figure} 
\plotone{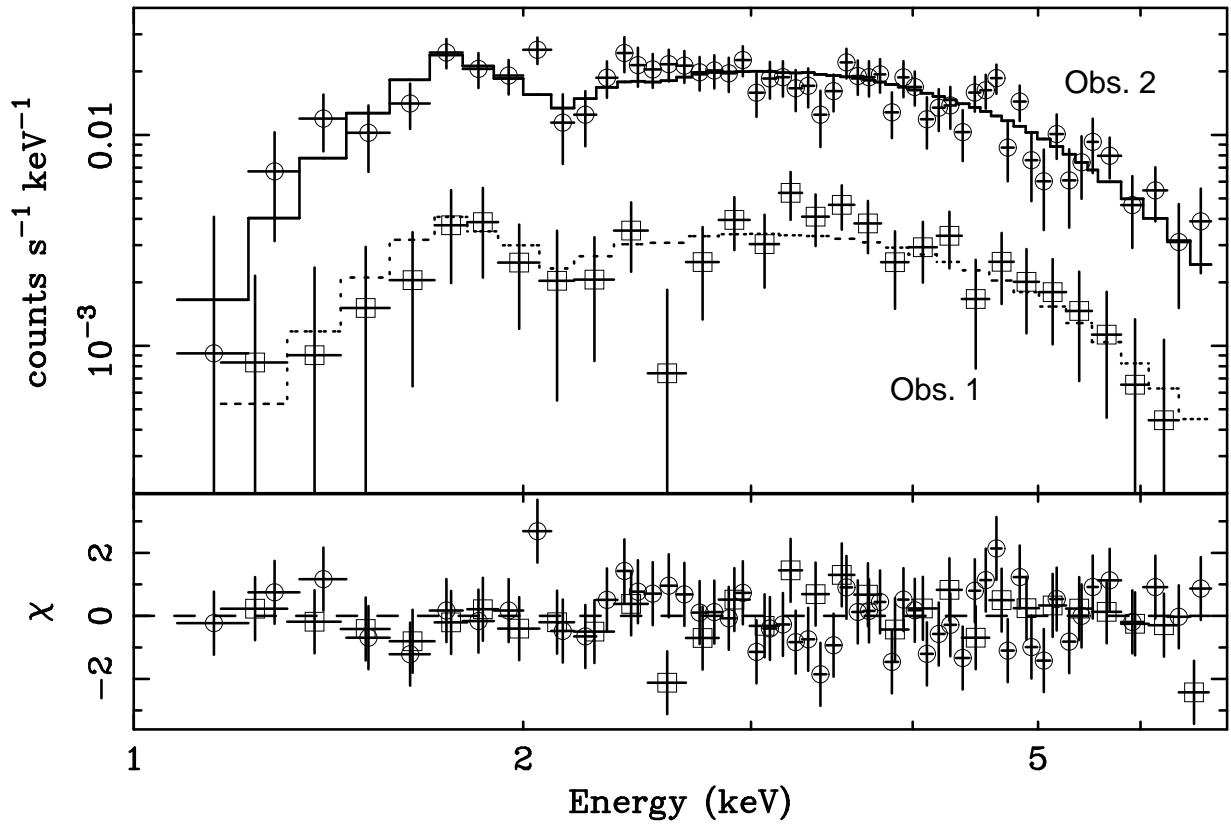}
\figcaption{The spectra of AX~J1843.8$-$0352 derived from Obs.\ 1 (squares) 
and 2 (circles). 
The best-fit power-law models in the simultaneous fitting are shown 
with dotted and solid lines for Obs.\ 1 and Obs.\ 2, respectively.
}
\end{figure}

\begin{figure} 
\plotone{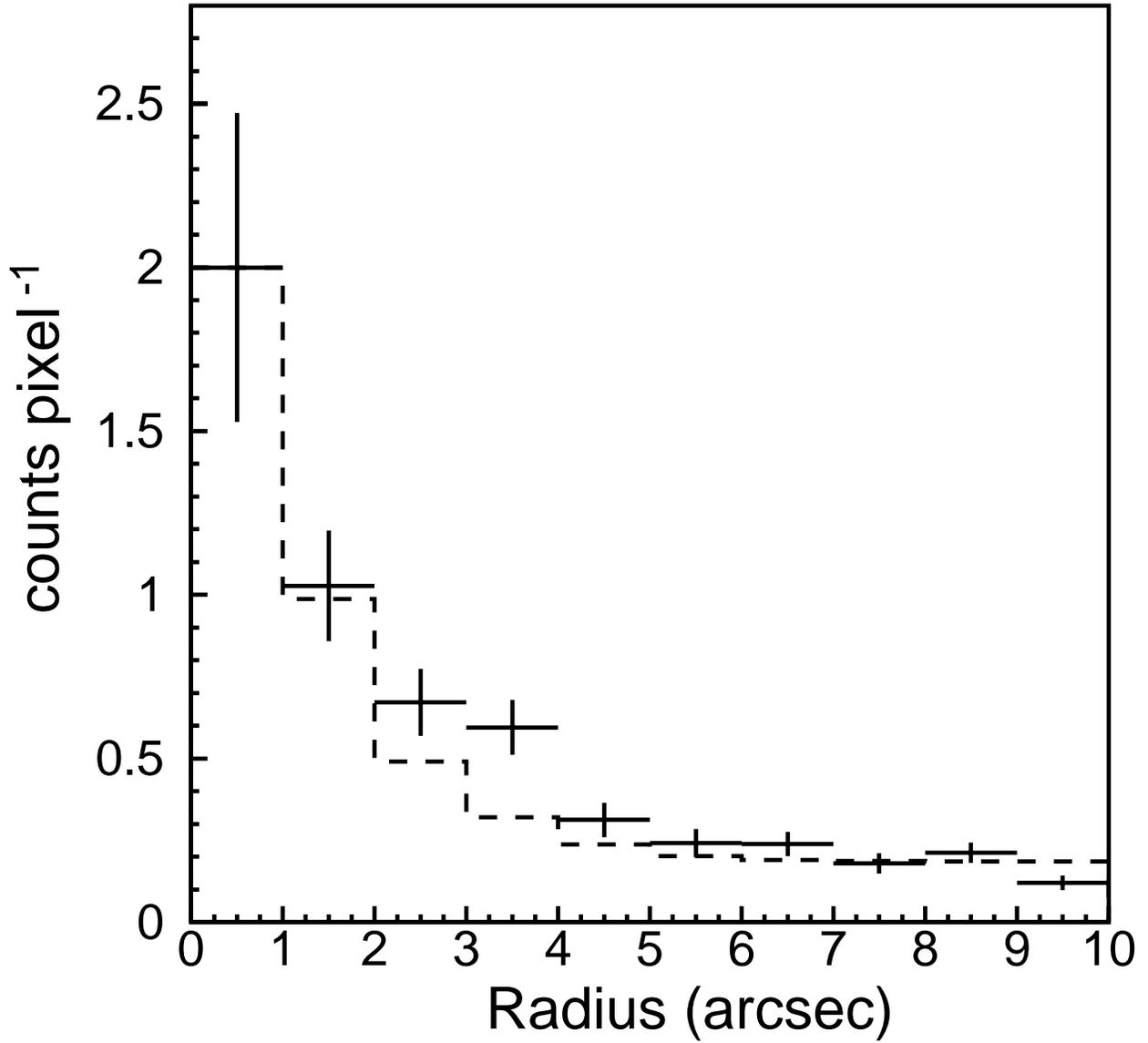}
\figcaption{The radial profile (1.0--6.0~keV) of 
the brightest point-like source in CXO~J184357$-$035441. 
One pixel is a square of $0\farcs 5 \times 0\farcs 5$.
The dashed line is the expected brightness by (PSF + constant). 
The PSF is normalized to the flux at the center.}
\end{figure}

\begin{figure} 
\plotone{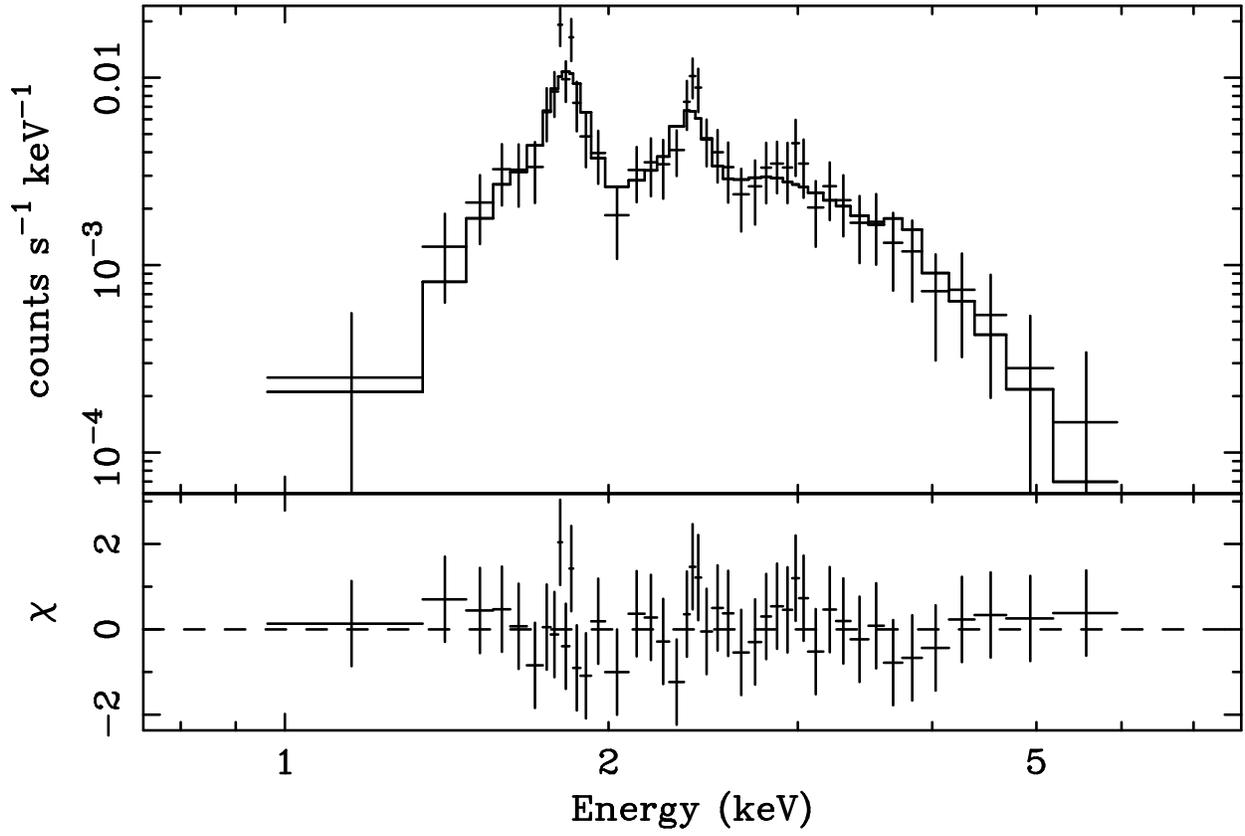}
\figcaption{The background-subtracted spectrum of CXO~J184357$-$035441. 
The best-fit NEI model is shown with the solid line.
}
\end{figure}

\clearpage

\begin{deluxetable}{lccccc}
\tablewidth{0pt}
\tablenum{1}
\tablecaption{Best-Fit Model Parameters for AX~J1843.8$-$0352}
\tablehead{
\colhead{}& \multicolumn{3}{c}{power-law}&\colhead{}&\colhead{NEI}\\
\cline{2-4}\cline{6-6}\\
\colhead{Parameter}&\colhead{Obs.\ 1}&\colhead{Obs.\ 2 }&\colhead{Obs.\ 1+2}&\colhead{}&\colhead{Obs.\ 1+2}
} 
\startdata
Photon index \dotfill &$2.4_{-0.9}^{+1.1}$ & $2.1_{-0.4}^{+0.4}$ & $2.1_{-0.3}^{+0.4}$ && \nodata \\
$kT[{\rm keV}]$\dotfill &\nodata & \nodata & \nodata && $5.4_{-1.6}^{+3.5}$ \\
Abundance\tablenotemark{a}\dotfill  & \nodata & \nodata & \nodata && $0.17_{-0.14}^{+0.17}$ \\
log($n_{\rm e} t$ [cm$^{-3}$~s]) \dotfill & \nodata & \nodata & \nodata && $11.1_{-0.7}^{+0.4}$ \\
$N_{\rm H}$~[$10^{22}$~cm$^{-2}$] \dotfill & $4.7_{-1.8}^{+2.8}$ &$3.7_{-0.6}^{+0.7}$ & $3.8_{-0.6}^{+0.7}$ && $3.5_{-0.5}^{+0.6}$ \\
Absorbed flux\tablenotemark{b}~[$10^{-14}$~erg~cm$^{-2}$~s$^{-1}$~arcmin$^{-2}$] \dotfill & 2.3 & 3.1 & 3.1\tablenotemark{c} && 3.0\tablenotemark{c}\\
Unabsorbed flux\tablenotemark{d}~[$10^{-14}$~erg~cm$^{-2}$~s$^{-1}$~arcmin$^{-2}$] \dotfill & 3.6 & 4.3 & 4.3\tablenotemark{c} && 4.1\tablenotemark{c} \\
$\chi ^{2}/$d.o.f. \dotfill &18.9/27&48.1/52&68.1/81&&65.3/79\\
\enddata
 \tablecomments{The errors correspond to 90\% confidence.}
 \tablenotetext{a}{Assuming the solar abundance ratio \citep{Anders1989}.}
 \tablenotetext{b}{Observed flux per unit area (arcmin$^{2}$) in 2.0--10.0 keV. }
 \tablenotetext{c}{The fluxes of ``Obs. 1+2'' are calculated from the 
normalizations for the spectrum of Obs.\ 2 which covers whole AX~J1843.8$-$0352. }
 \tablenotetext{d}{Absorption-corrected flux per unit area (arcmin$^{2}$) in 2.0--10.0 keV.}
\end{deluxetable}

\begin{deluxetable}{lc}
\tablewidth{0pt}
\tablenum{2}
\tablecaption{Best-Fit Parameters for CXO~J184357$-$035441 by the NEI Model}
\tablehead{\colhead{Parameter}&\colhead{Value}} 
\startdata
$kT$[keV] \dotfill & $0.71_{-0.16}^{+0.22}$ \\
Abundance\tablenotemark{a} \dotfill &  $1.1_{-0.7}^{+1.4}$    \\
log($n_{\rm e} t$~[cm$^{-3}$~s]) \dotfill & $10.4_{-0.3}^{+1.5}$   \\
$N_{\rm H}$[$10^{22}~{\rm cm}^{-2}$] \dotfill & $7.4_{-1.2}^{+1.6}$     \\
Absorbed flux\tablenotemark{b}~[erg~cm$^{-2}$~s$^{-1}$] \dotfill & $1.1 \times 10^{-13}$ \\
Unabsorbed flux\tablenotemark{c}~[erg~cm$^{-2}$~s$^{-1}$] \dotfill & $4.5 \times 10^{-11}$ \\
$\chi ^{2}/$d.o.f.& 22.2/39\\
\enddata
 \tablecomments{The errors correspond to 90\% confidence.}
 \tablenotetext{a}{Assuming the solar abundance ratio \citep{Anders1989}.}
 \tablenotetext{b}{Observed flux in 0.7--10.0 keV.}
 \tablenotetext{c}{Absorption-corrected flux in 0.7--10.0 keV.}
\end{deluxetable}

\begin{deluxetable}{lcc}
\tablewidth{0pt}
\tablenum{3}
\tablecaption{Best-Fit Parameters by the SRCUT Model}
\tablehead{\colhead{Parameter}&\colhead{$\alpha =0.5$$^{a}$}&\colhead{$\alpha =0.6$$^{a}$}} 
\startdata
Cut-off frequency~[$10^{17}$~Hz] \dotfill & $7.2 (2.8-27)$& $11(2.2-92)$\\
Flux density$^{b}$~[Jy] \dotfill & $1.8 (1.1-4.6) \times 10^{-2}$&$0.11 (0.07-0.45)$    \\
$N_{\rm H}$[$10^{22}~{\rm cm}^{-2}$] \dotfill & $3.6_{-0.2}^{+0.3}$ &$3.7_{-0.3}^{+0.2}$    \\
\hline
$B_{\rm eq}$~[$\mu$G]  \dotfill & 8 & 6\\
$E_{\rm eq}$~[erg] \dotfill & $3 \times 10^{47}$&$2 \times 10^{47}$\\
\enddata
 \tablecomments{The errors and the error ranges correspond to 90\% confidence.}
 \tablenotetext{a}{$\alpha$ was fixed to the value.}
 \tablenotetext{b}{Flux density at 1.5~GHz.}  
\end{deluxetable}

\end{document}